\documentclass[english, 12pt]{elsart}

\usepackage[T1]{fontenc}
\usepackage{babel}
\usepackage{amsmath}
\usepackage{graphics}

\newcommand\lsim{\mathrel{\rlap{\lower4pt\hbox{\hskip1pt$\sim$}}
    \raise1pt\hbox{$<$}}}
\newcommand\gsim{\mathrel{\rlap{\lower4pt\hbox{\hskip1pt$\sim$}}
    \raise1pt\hbox{$>$}}}

\begin{document}
\begin{frontmatter}

\title{On the Behaviour and Stability of Superconducting Currents}

\author{Y. Lemperiere\thanksref{YFJ}}
\and
\author{E.P.S. Shellard\thanksref{EPS}}
\address{DAMTP, University of Cambridge, Wilberforce Road, Cambridge 
CB3 0WA, UK}
\thanks[YFJ]{Electronic address: Y.F.J.Lemperiere$\,$\hbox{\rm
@}$\,$damtp.cam.ac.uk}
\thanks[EPS]{Electronic address: E.P.S.Shellard$\,$\hbox{\rm @}$\,$damtp.cam.ac.uk}

\begin{abstract}
: We present analytic and numerical results for the evolution of currents on
superconducting strings in the classical $U(1) \times U(1)$ model. 
We derive an energy functional for the currents and charges on these strings,
establishing rigorously that minima
should exist in this model for loops of finite size
(vortons) if both charge and current are present on the worldsheet. We then study the stability of the currents on these
strings, and we find an analytic criterion for the onset of
instability (in the neutral limit).  This limit specifies a lower
maximal current than previous heuristic estimates. We conclude with a
discussion of the evolution of loops towards their final vorton state in 
the model under consideration.
\end{abstract}


\end{frontmatter}


SLAC: hep-ph/0207199

\section{Introduction}

Topological defects are a class of exact solutions in field theories
whose stability is enforced by topological reasons. In particular,
strings, the class of defects associated with a non-trivial first
homotopy group of the vacuum manifold, have been widely studied, since
they seem to appear in a variety of generalisations of the Standard
Model (GUTs, SUSY etc.). They are for example associated with the
spontaneous symmetry breakdown of a $U(1)$ symmetry, like the
axion or baryon symmetry (global), or electromagnetism (local) in
superconductivity. Strings could therefore appear during a
cosmological phase transition, and are prime candidates for a number of
astrophysical puzzles, like the dark matter of the universe or the
origin of the most energetic cosmic rays (for a review
of cosmic defects, see ref.~\cite{VS:CS}). 

In this paper, we shall be interested in a class of
defects where the string's field is coupled to another scalar field,
which allows the build up of charge and currents on the
worldsheet. Because of the non-dissipative properties of the currents on these 
strings, they are called superconducting.

The plan of this paper is as follows: in the next section, we will
discuss the Lagrangian under consideration, and show how the amount of
charge and currents is limited; in the following part, we will carry out
an analysis of the energy functional of the condensate, to get a very
simple analytic expression for it. We will then use our results to
discuss the possibility of forming stable loops of 
superconducting strings. Finally, we will study the stability of
currents on the string's worldsheet and derive an exact analytic
result for the onset of instability, which we shall illustrate by
numerical simulations obtained from our full 3D field theory code.

\section{The field theory model}

The model under consideration here is the original $U(1)\times U(1)$
model, first proposed by Witten \cite{WI:SCS}, and based on the Lagrangian
\begin{eqnarray}
\mathcal{L} &= (\partial_{\mu}\phi)(\partial^{\mu}\phi)^+ +
(\partial_{\mu}\sigma)(\partial^{\mu}\sigma)^+ -
\frac{\lambda_\phi}{4}(|\phi|^2-\eta_\phi^2)^2 \nonumber\\
&~~~~~~~~~~- \frac{\lambda_\sigma}{4}(|\sigma|^2-\eta_\sigma^2)^2 -
\beta|\phi|^2|\sigma|^2 \,.
\label{lagrangian}
\end{eqnarray}
where $\phi$ and $\sigma$ are complex scalar fields and $\lambda_\phi,
\,\lambda_\sigma,\,\eta_\phi,\,\eta_\sigma$ and $ \beta$ are positive 
constants.  We can arrange the parameters in this model such that the
ground state has the $\phi$-symmetry broken ($|\phi| =\eta_\phi\ne0$), 
while $\sigma$ remains symmetric $(|\sigma|=0)$.  Under these 
circumstances, vortex solutions exist in the $\phi$-field. Here, we
shall assume we have a vortex-string lying along the $z$-axis; in 
azimuthal coordinates, this solution takes the form, 
\begin{equation}
\phi(r,\theta,z) = |\phi|(r) e^{i\theta}\,,
\label{vortexsolution}
\end{equation}
with $|\phi|(0)=0$ at the vortex centre and $|\phi|\rightarrow \eta_\phi$ 
as $r\rightarrow \infty$.

We wish to consider the conditions under which a condensate in the 
$\sigma$-field can emerge in the core of the string.  This condensate
can also carry currents and charges along the string, so we will represent 
it by the following ansatz which describes the dependence of these
excitations on $z$ and $t$:
\begin{equation}
\sigma = |\sigma|(r)\exp^{i(\omega t +kz)},
\label{field}
\end{equation}
We can see that the presence of charge and current causes 
a change in the Lagrangian $\delta\mathcal{L} =
(\omega^2 - k^2)|\sigma|^2$, which alters the mass term
\cite{DS:SCS1,DS:SCS2}, and the
vacuum expectation value of $\sigma$; indeed, we can rewrite the
Mexican hat potential for $|\sigma|$ as:
\begin{equation}
V_{\sigma} = \frac{\lambda_{\sigma}}{4}\left[|\sigma|^2 -
\left(\eta_{\sigma}^2 + \frac{2}{\lambda_{\sigma}}(\omega^2 -
k^2)\right)\right]^2 .
\label{newpot}
\end{equation}

From the expression (\ref{newpot}), we 
see that the $\phi$-symmetry will remain broken
in the vacuum, with $|\phi| = \eta_{\phi}$,  as long as we satisfy
\begin{equation}
\lambda_{\phi}\eta_{\phi}^4 > \lambda_{\sigma}\left(\eta_{\sigma}^2 +
\frac{2(\omega^2 - k^2)}{\lambda_{\sigma}}\right)^2  ,
\label{c1}
\end{equation}
If we want the $\sigma$-symmetry to be remain unbroken in the 
ground state ($|\sigma|=0$), then we need to ensure that the mass 
term for $\sigma$
is positive at infinity, that is, 
\begin{equation}
m_{\sigma}^2 \equiv \beta\eta_{\phi}^2 -
\frac{1}{2}\lambda_{\sigma}\eta_{\sigma}^2 - \omega^2 + k^2 >0 \,.
\label{c2}
\end{equation}
Now at the centre of the string we have $|\phi|=0$, so this mass
term can become negative and $\sigma$ can develop a 
non-vanishing expectation value, provided that
\begin{equation}
\frac{1}{2}\lambda_{\sigma}\eta_{\sigma}^2 + \omega^2 - k^2 >0 \,.
\label{c2prime}
\end{equation}

However, the condition (\ref{c2prime}) is not sufficient to 
obtain a $\sigma$-condensate within the string, since we also have to
consider its gradient energy cost. To determine this we follow the
analysis of Haws \emph{et al.} \cite{HHT:SCS} and study the stability of the
trivial solution $\sigma = 0$ with perturbations of the form $\sigma =
|\sigma(x, y)|e^{i\nu t}$, with $|\sigma|<<1$. The field equation,
using the modified Mexican Hat potential (\ref{newpot}), then becomes
\begin{equation}
-\partial_r\partial^r\sigma - \left(\frac{1}{2}\lambda_{\sigma}[\eta_{\sigma}^2 +
\frac{2(\omega^2 -
k^2)}{\lambda_{\sigma}}] - 
\beta\lambda_{\phi}\eta_{\phi}^4r^2\right)\sigma = \nu^2\sigma \, ,
\label{HO}
\end{equation}
which is an eigenvalue equation with a harmonic oscillator-like
potential. Hence, we know the value of the ground state eigenvalue $\nu_0$, and
we see that the trivial solution will be unstable for $\nu_0^2 < 0$ or,
equivalently,
\begin{equation}
k^2 - \omega^2 < \frac{1}{2}\lambda_{\sigma}\eta_{\sigma}^2 -
\sqrt{\beta\lambda_{\phi}}\eta_{\phi}^2 \,.
\label{c3}
\end{equation}

\begin{figure}
\resizebox{13cm}{6.5cm}{\includegraphics{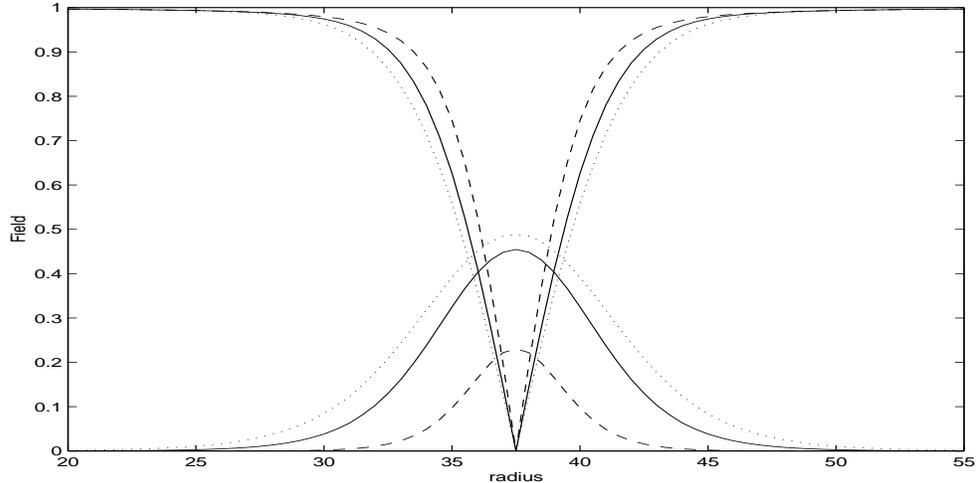}}
\caption{Profiles of the vortex and the condensate fields, with no
charge and current (solid lines). The effects of adding a significant
 current (dashed) or charge density
(dotted) is also illustrated; the (anti-)quenching effect of 
current (charge) is clear.}
\label{profiles}
\end{figure}

If (\ref{c1}), (\ref{c2}), and (\ref{c3}) are satisfied, $\sigma$ will
form a condensate of width $\delta_{\sigma} \simeq m_{\sigma}^{-1}$, with 
two conserved quantities, the usual Noether
charge $Q$ and a topological charge $N$, where:
\begin{equation}
Q = \omega \int dz \int dS\, |\sigma|^2 \, , \qquad\qquad N = \int
dz\, \frac{k}{2\pi} \, ,
\label{charges}
\end{equation}
and a current flowing along the $z$ direction:
\begin{equation}
J_z = k \int dz \int dS\, |\sigma|^2 \, .
\label{curernt}
\end{equation}
The field configuration for a vortex with a condensate can be seen in
 fig.~\ref{profiles},
with the influence of $k$ and $\omega$ also shown.

To conclude this section, we note that in the bare case ($\omega = k =
0$), (\ref{c1}), (\ref{c2}), and (\ref{c3}) lead to the following
inequalities on the parameters of the theory:
\begin{equation}
\beta < \frac{\lambda_{\sigma}}{4} \, , \qquad 2\eta_{\sigma}^2 <
\eta_{\phi}^2 \, .
\label{param}
\end{equation}
These equations will prove to be useful in the rest of the paper.

\section{The energy functional}

Starting from the Lagrangian (\ref{lagrangian}), we can easily switch to a
Hamiltonian formalism. In particular, we will be interested in the
following in the energy of the condensate, which can be written in
the form:
\begin{equation}
\mathcal{E}_{\sigma} = \int dz \int dS \left( |\nabla_r\sigma|^2 + (k^2 +
\omega^2)|\sigma|^2 + (\beta |\phi|^2 -
\frac{1}{2}\lambda_{\sigma}\eta_{\sigma}^2)|\sigma|^2 +
\frac{\lambda_{\sigma}}{4}|\sigma|^4 \right) \,.
\label{ham}
\end{equation}
This is a rather complicated expression, so in order to simplify it
we consider the equation of motion for $\sigma$:
\begin{equation}
-\partial_r\partial^r\sigma + (k^2 - \omega^2)\sigma + (\beta|\phi|^2
- \frac{1}{2}\lambda_{\sigma}\eta_{\sigma}^2)\sigma +
\frac{\lambda_{\sigma}}{2}|\sigma|^2\sigma = 0 \, .
\label{eom}
\end{equation}
Multiplying (\ref{eom}) by the complex conjugate
$\sigma^+$, integrating by parts, and
then inserting the result in (\ref{ham}), we find:
\begin{equation}
\mathcal{E}_{\sigma} = -\frac{\lambda_{\sigma}}{4}\int d^3r\, |\sigma|^4 +
2 \int d^3r\, \omega^2 |\sigma|^2 \, .
\label{nham}
\end{equation}
(It is easy to generalise this equation to the gauged case, making the
obvious replacement $\omega \to \omega - A_t$.)

Now, if we recall the expression for the conserved charge $Q$ (\ref{charges}),
we can rewrite the energy as:
\begin{equation}
\mathcal{E}_{\sigma} = -\frac{\lambda_{\sigma}}{4}\int d^3r |\sigma|^4 +
\frac{2Q^2}{\int d^3r\, |\sigma|^2} \, .
\label{nnham}
\end{equation}
This equation (\ref{nnham}) is interesting because it expresses the
energy in the condensate as a function of conserved quantities, and of
the two integrated moments $\Sigma_2$ and $\Sigma_4$, where
\begin{equation}
\Sigma_{n} = \int dS\, |\sigma|^n \, .
\label{Sigma}
\end{equation}

Note that $d\Sigma_n / dz = 0$ with the ansatz (\ref{field}), and so
we can carry out the $z$-integration in (\ref{nnham}) for a segment of
finite length $L$ to obtain
\begin{equation}
\mathcal{E}_{\sigma} = -\frac{\lambda_{\sigma}}{4} \Sigma_4 L +
\frac{2Q^2}{\Sigma_2 L} \, ,
\label{newham}
\end{equation}
which is a purely analytic expression.

\begin{figure}
\resizebox{11cm}{7cm}{\includegraphics{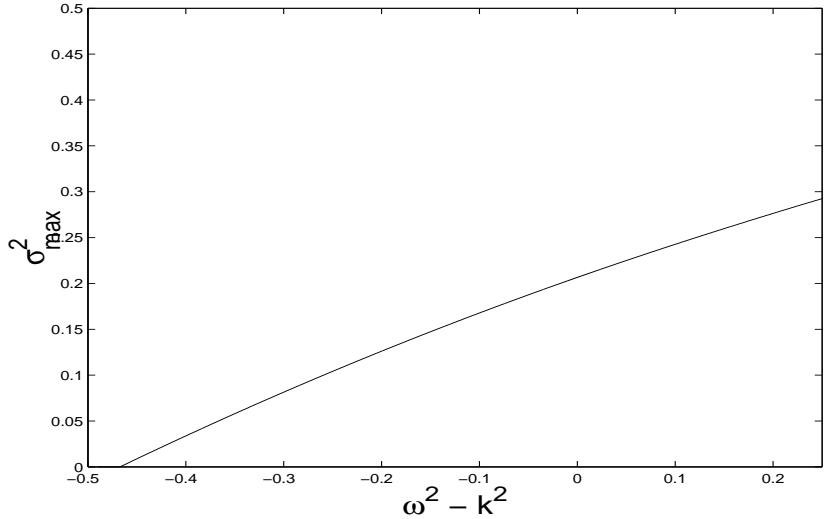}}
\caption{Plot of the numerical values for $\sigma_{\rm max}^2$, as a
function of $\omega^2 - n^2$.}
\label{sigma2}
\end{figure}

\begin{figure}
\resizebox{13cm}{6.5cm}{\includegraphics{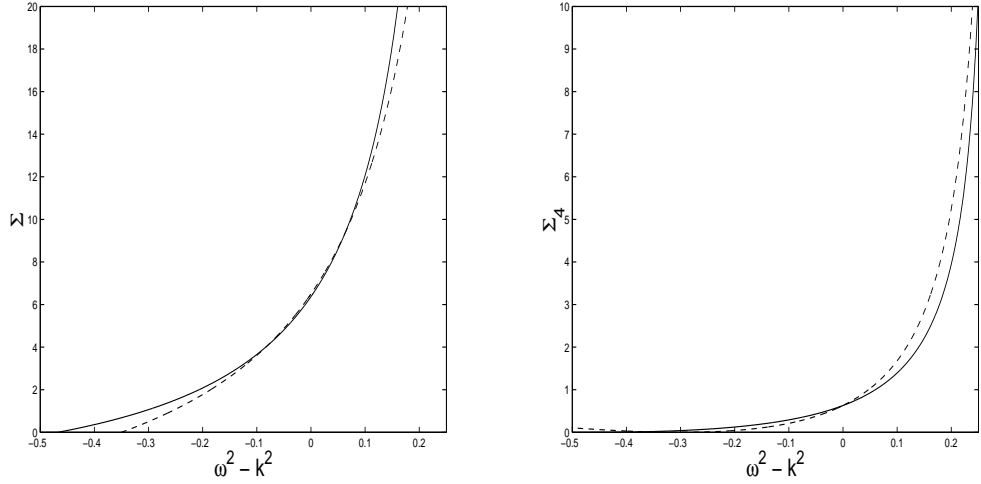}}
\caption{Plots of the numerical values for $\Sigma_2$ and $\Sigma_4$
(plain), and their analytic estimates (dashed) from (\ref{Sigmaf}).}
\label{Sigmas}
\end{figure}

This is the form of the energy functional we sought. All
that is now required is to find an ansatz for the $\Sigma_n$, which can be
done by noticing that:
\begin{equation}
\Sigma_{n} = \int dS\, |\sigma|^n ~~\simeq~~ \sigma_{\rm max}^n \times
\delta_{\sigma}^2 ~~\simeq~~ \frac{\sigma_{\rm max}^n}{m_{\sigma}^2} \, ,
\label{Sigmaan}
\end{equation}
where $\delta_{\sigma}$ is the average width of the condensate, and
$\sigma_{\rm max}$ is its maximum height in the vortex core. 
Now, we have the mass 
given by 
\begin{equation}
m_{\sigma}^2 = \beta\eta_{\phi}^2 -
\frac{1}{2}\lambda_{\sigma}\eta_{\sigma}^2 - \omega^2 + k^2 \equiv
\omega_c^2 - \omega^2 + k^2\,,
\end{equation}
and we can see from fig.~\ref{sigma2} 
that, up to
a very good approximation, we can take 
\begin{equation}
\sigma_{\rm max}^2 \equiv \sigma_0(k_c^2
- k^2 + \omega^2)\,,
\end{equation}
where $k_c^2 \simeq
\frac{1}{2}\lambda_{\sigma}\eta_{\sigma}^2 -
\sqrt{\beta\lambda_{\phi}}\eta_{\phi}^2$ is the 
critical winding number density at which the condensate 
must vanish (obtained from (\ref{c3})).  (We note
that in the
gauged case, the picture is qualitatively the same, but that the gauge 
fields maintain the condensate against the
quenching effect of the current; the height is then 
nearly constant in a somewhat wider range of $\omega^2 -
k^2$. Ultimately, however, $k_c^2$ is the same for both 
gauged and global cases.) 
If we further assume that $k_c \simeq \omega_c \simeq m_o$, 
we can factorize it out in
(\ref{Sigmaan}) to yield:
\begin{equation}
\Sigma_{n} = \Sigma_{no}\frac{(1+v)^{\frac{n}{2}}}{1-v} \, ,
\label{Sigmaf}
\end{equation}
with $v = (\omega^2 - k^2)/{m_o^2}$ and $\Sigma_{no} = \Sigma_n
(\omega^2 - k^2 = 0)$ is the value of $\Sigma_n$ in the so-called
\emph{chiral case} ($\omega=\pm k$). In particular, for
$n = 2$ this reduces to:
\begin{equation}
\Sigma ~\equiv~ \Sigma_2 ~=~ \Sigma_o \,\frac{m_o^2 +(\omega^2 - k^2)}{m_o^2 -
(\omega^2 - k^2)} \, .
\end{equation}
As can be seen from
fig.~\ref{Sigmas}, the agreement between the numerical calculation
and our analytic estimate (\ref{Sigmaf}) is remarkably good. 

\section{The behaviour of superconducting loops}

Cosmic strings tend to intercommute, and therefore we can expect the copious
production of loops from a
network of strings. Superconducting
strings follow the same behaviour \cite{LM:SCS}, but this time 
the charge and current due to the additional structure
may prevent the shrinkage of the loop and stabilise the
configuration \cite{DS:SCS2}. These stable loops or 
\emph{vortons}, effectively held up by their angular momentum, have
been the subject of numerous studies (see, for example, refs.~\cite{DS:CV,MS:VF,CD:CV,CP:VO,BPOV:VO}).

Using the results of the previous sections, we would like 
to study the behaviour of these loops, starting from our analytic
result for the energy of the condensate (\ref{newham}). However, 
we must also include the energy in the vortex, which we will model by
$\mathcal{E}_{\phi} = \mu L$, where $\mu \sim \mathcal{O} (\eta_\phi^2)$ 
is the mass per unit length
of the string.
We can then write our energy functional for the loop as
\begin{equation}
\mathcal{E} = \mu L - \frac{\lambda_{\sigma}}{4}\Sigma_4 L +
\frac{2Q^2}{\Sigma_2 L} \, .
\label{entot}
\end{equation}
We will now discuss the various possible regimes for the charges
and currents on these loops:

\medskip
\noindent \emph{The chiral case, $\omega^2 = k^2$:}

\nopagebreak
\noindent We will first consider 
this limit, where $\omega^2 - k^2 =
0$ (see refs.~\cite{DKPS:CCS,PD:CCS} for a study of chiral loops using
the 2D approach of \cite{CP:VO}). In this regime, we always have $\Sigma = \Sigma_o$ and
$\Sigma_4 = \Sigma_{4o}$. Since these values remain fixed as the loop 
shrinks, it is easy to
vary (\ref{entot}) to see that it will have a minimum for 
\begin{equation}
L^2 = \frac{2Q^2}{\Sigma_o(\mu -
\frac{\lambda_{\sigma}}{4}\Sigma_{40})} = \frac{2N^2 \Sigma_o}{(\mu -
\frac{\lambda_{\sigma}}{4}\Sigma_{40})} \, .
\label{chiral}
\end{equation}
Hence, the loop will contract (or expand) 
until $\omega^2 = k^2 = {(\mu -
\frac{1}{4}\lambda_{\sigma}\Sigma_{4o})}/({2 \Sigma_o})$.

For this to be well defined, we should verify that
$\mathcal{E}_{k = \omega = 0} = \mu -
\frac{\lambda_{\sigma}}{4}\Sigma_{4o}$ is positive. To do so, we 
note that this is the total energy in the bare superconducting string, and
that the profiles $|\phi| (r)$ and $|\sigma| (r)$ minimise the value of
this functional. By considering $\phi (\lambda r)$ and $\sigma (\lambda
r)$, differentiating the field energy functional (\ref{ham}) with
respect to $\lambda$ and setting this quantity to $0$ at $\lambda =
1$, we obtain the result:
\begin{equation}
\mathcal{E}_{k = \omega = 0} = \int d^3r
(|\nabla\phi|^2 + |\nabla\sigma|^2) > 0 \,.
\label{chipos}
\end{equation}
This ensures that the loop length $L^2$ is indeed well defined and that chiral
loops must have a stable minimum of their energy. (Of course, we assume 
here that chiral currents are long-lived 
and that the final stable loop size is significantly larger than the vortex
width.)

\medskip
\noindent\emph{The electric and magnetic cases, $\omega^2 - k^2 \not= 0$:}
\nopagebreak

\noindent To study these, we have to
manipulate the form of the loop energy. Using the fact that $N = k L$ 
and $Q = \omega \Sigma L$ are
conserved during the evolution of the loop, we can express $L$ in the
following terms:
\begin{equation}
L^2 = \frac{N^2}{\omega^2 - k^2}\left(\frac{\Sigma_{QN}^2}{\Sigma^2} - 1\right)
\, ,
\label{L2}
\end{equation}
where $\Sigma_{QN} = Q/N$. (Note that the ratio $Q/N$ is defined at 
loop creation and is assumed to be conserved thereafter.) 
By substituting this back in (\ref{entot}),
we get an expression for the energy as a function of
$\omega^2 - k^2$:
\begin{equation}
\mathcal{E} =N \left[ \frac{(\mu  -
\frac{\lambda_{\sigma}}{4}\Sigma_4)}{\Sigma}
\sqrt{\frac{\Sigma_{QN}^2-\Sigma^2}{\omega^2-k^2}} + 2
\Sigma_{QN}^2 \sqrt{\frac{\omega^2-k^2}{\Sigma_{QN}^2-\Sigma^2}} \right] \,
.
\label{nentot}
\end{equation}
Here, note that $\Sigma_n$ is clearly a function of $\omega^2 - k^2$, since
it is determined by the solution of the field equations,
which only involve $\omega^2 - k^2$.

It is also important to notice that $\Sigma_{QN}$ can be expressed
as: $\Sigma_{QN} = \frac{\omega}{k}\Sigma$. So, using the
terminology of Carter and Peter \cite{CP:VO}, in the electric case 
($\omega^2 > k^2$),
we have $\Sigma_o < \Sigma < \Sigma_{QN}$, while in the
magnetic case ($k^2 > \omega^2$), we have $\Sigma_{QN} < \Sigma <
\Sigma_o$. This implies that for each loop, $\Sigma$ can only 
take a limited range
of values, the bounds of which are fixed, on the one hand, by the 
parameters in the Lagrangian determining the bare $\Sigma_o$ and,
on the other, by the amount of charge and current on the loop
fixing $\Sigma_{QN}$. 

A simple analysis of (\ref{nentot}) now tells us that, as $\Sigma^2
\to \Sigma_{QN}^2$ (which is achieved for a non-vanishing value of
$\omega^2 - k^2$), the energy $\mathcal{E}$ goes to $+\infty$, 
while in the opposite limit as $\omega^2 - k^2 \to 0$ then $\mathcal{E}$
diverges once again, that is,  provided we have
$\mu - \frac{\lambda_{\sigma}}{4}\Sigma_4 > 0$. Hence, as long
as this condition is satisfied, then the loop energy 
$\mathcal{E}$ must have a minimum, an equilibrium state 
corresponding to the vorton.

\begin{figure}
\resizebox{13cm}{6.5cm}{\includegraphics{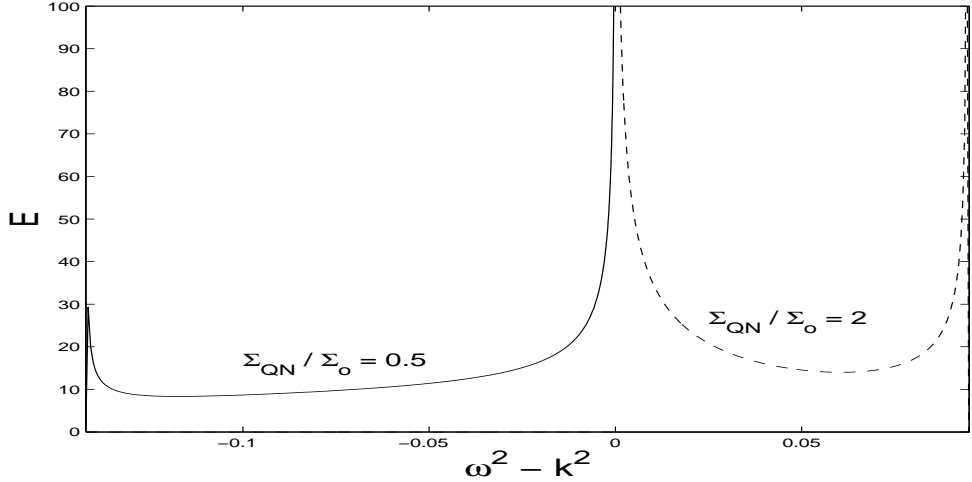}}
\caption{This figure shows the energy E as a function of $\omega^2 -
k^2$, in the mangetic regime ($\Sigma_{QN} / \Sigma_o = 0.5$ on the left), and in the
electric regime ($\Sigma_{QN} / \Sigma_o = 2$ on the right).}
\label{Energy}
\end{figure}

As can be seen from fig.~\ref{Energy}, the vorton state is usually reached
when $\Sigma \simeq \Sigma_{QN}$, which means that $\omega /k
\simeq 1$.  This appears to justify the assertion that 
$\frac{\omega}{k} \to 1$ is an 
attractor \cite{DS:SCS2}, but it does {\it not} imply that the 
chiral state with $\omega^2-k^2 =0$ is an attractor.  On the contrary,
a typical loop created with small charges and currents ($|\omega_{\rm i}^2-
k_{\rm i}^2|
\hbox{$<\!<$} m_o^2$) will contract to a final vorton state which 
is strongly in the magnetic or electric regimes with $|\omega^2-k^2|
\simeq {\mathcal O}(m_o^2)$.  Even if the initial state was fine-tuned very 
close to the chiral case $\omega_{\rm i}^2\simeq
k_{\rm i}^2$, then the final state always will be less chiral.  

The last point is to check the required condition that $\mu^2 -
\frac{\lambda_{\sigma}}{4}\Sigma_4 > 0$ when $\omega^2 - k^2 \simeq
0$. But in this case, $\mu$ and $\Sigma_4$ tend to their chiral limits, and we
have already proved in eq.~(\ref{chipos}) that in this limit, this
quantity is 
positive,
$\mathcal{E}_{k = \omega = 0} > 0$ .
So this in turn ensures
that the condition for the existence of a vorton holds for the whole
magnetic and electric
regime, except in the limiting cases where $\Sigma_{QN} = 0$ or
$\Sigma_{QN} = \infty$.

\medskip
\noindent\emph{Limiting cases $Q = 0$ and $N = 0$:} 
\nopagebreak

\noindent Let us first consider the
first of these cases, the so-called \emph{spring}
\cite{HHT:SCS,CTM:DSC}, with no charge on the loop. It is possible to
study analytically the behaviour of these objects in this model, if
one further assumes that there is no backreaction of the condensate on
the vortex field. This is a legitimate first approximation, if we
consider the profiles shown on fig.~\ref{profiles}. Then, with our
ansatze (\ref{Sigmaf}), we are able
to differentiate (\ref{nentot}) with respect to $k$ (note that
$\Sigma_{QN} = 0$), from which we obtain a polynomial
in $v = -k^2/k_c^2$:
\begin{eqnarray}
P(v) \equiv \frac{(1-v)^2}{N}\frac{d{\mathcal E}}{dk} &= &\frac{\lambda_{\sigma}}{4}\Sigma_{4o}v^3 - (\mu +
\frac{5\lambda_{\sigma}}{4}\Sigma_{4o})v^2 \nonumber\\
&&+ (\mu -
\frac{5\lambda_{\sigma}}{4}\Sigma_{4o})v -\mu +
\frac{\lambda_{\sigma}}{4}\Sigma_{4o} \, .
\label{derivate}
\end{eqnarray}

\begin{figure}
\resizebox{13cm}{6.5cm}{\includegraphics{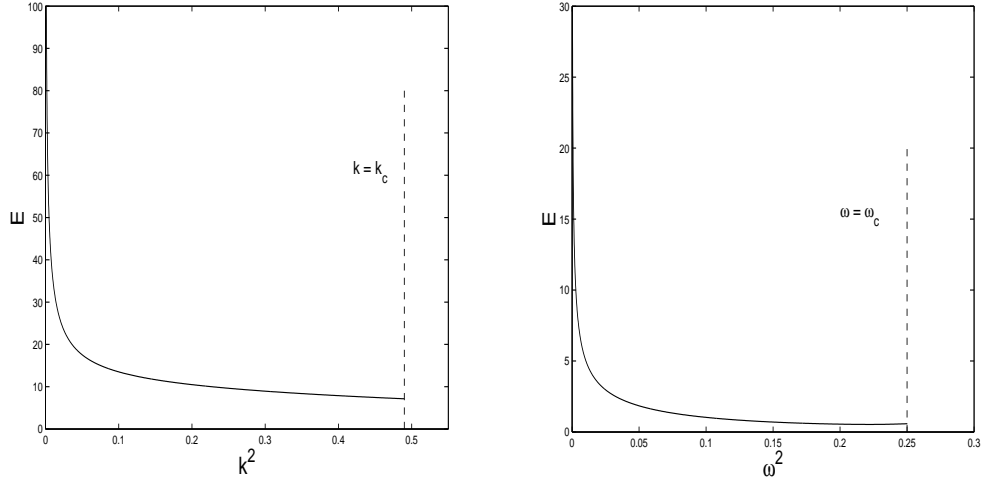}}
\caption{This figure shows the behaviour of the energy functional for
the spring ($E$ is plotted as a function of $k^2$) and the Q-loop ($E$
is then a function of $\omega^2$). It seems that no stable
configuration of finite radius can be expected from these
configurations.}
\label{Elim}
\end{figure}

No springs will form if $P$ is always negative.  
Since we are restricted to the interval $-1<v<0$
and as both $P(0)<0$ and $P(-1)<0$, this will be the case 
if the derivative $P'$ never
vanishes between $0$ and $-1$.  Now $P'$ is a second
order polynomial, and some algebra shows that it has a negative
minimum for $v > 0$. Thus, it is enough to check that $P'(0) =
2\mu - \frac{5\lambda_{\sigma}}{4}\Sigma_{4o} > 0$ (it can be seen
that $P'(-1) >0$ by direct computation). To prove this, we
note that:
\begin{equation}
\frac{\lambda_{\sigma}}{4}\Sigma_{4o} < \frac{\lambda_{\sigma}}{4}
\frac{\eta_{\sigma}^4}{m_o^2} \lsim \eta_{\sigma}^2 <
\frac{\eta_{\phi}^2}{2} \simeq \frac{\mu}{4\pi} \, ,
\label{long}
\end{equation}
where we have assumed that $\frac{\lambda_{\sigma}}{2}
\frac{\eta_{\sigma}^2}{m_o^2} \lsim 2$, and that the string energy density is not too
different from its critical coupling value. This indicates that for
reasonable parameters, spring formation is excluded. The
behaviour of the energy functional is shown in fig.~\ref{Elim}.
Of course, for the gauged case it is possible that additional terms
from magnetic pressure could stabilise the configuration from collapse
\cite{HHT:SCS}.
However, since these are logarithmic corrections, such gauged
springs could only exist on astrophysical scales.

We have also studied the energetic behaviour of the Q-loop, with $N =
0$. In this case, we can rewrite the energy functional as:
\begin{equation}
\mathcal{E}_Q = (\mu -
\frac{\lambda_{\sigma}}{4}\Sigma_4)\frac{Q}{\omega \Sigma} + 2\omega Q
\, .
\label{Qenergy}
\end{equation}
The typical behaviour of this functional as $\omega$ varies is shown
on fig.~\ref{Elim}. As can be seen, there does not appear to be any stable
configuration, since the minimum of the energy is for $\omega =
\omega_c$, that is for $L = 0$.

\section{The stability of the superconducting current}

We have seen in the previous section that the model derived from
(\ref{lagrangian}) seems to lead to stable loop solutions, but the
issue of classical stability has not yet been addressed in a full
analytic treatment (for
the quantum stability, refer to \cite{BPOV:SCS}).  This analysis is 
inspired in part by a heuristic argument in ref.~\cite{TH:SCS} 
suggesting that classical instabilities could develop in currents with 
sufficiently high winding. Because of the Lorentz
invariance of the theory, we have only 3 cases to consider: pure
current ($\omega = 0$), pure charge ($k = 0$), and the chiral case
$\omega = k$. In this section, we wish to study the perturbations around the
pure
current solution. To proceed, we decompose our field in
the following way: $\sigma = S_k(r)(e^{ikz} + \nu(z))$, where $S_k$ is
the radial part of the unperturbed solution and $\nu$ is an arbitrary
perturbation.

The change in $\mathcal{E}_{\sigma}$ is easily computed:
\begin{equation}
\delta^2 \mathcal{E}_{\sigma} = \int dz \left(\Sigma_2 |\nabla_z
\nu|^2 + \gamma |\nu|^2 + \frac{\lambda_{\sigma}}{4}\Sigma_4 [e^{-ikz}\nu +
e^{ikz}\nu^+]^2 \right) \, ,
\label{delta}
\end{equation}
where 
\begin{equation}
\gamma = \int dS \, (|\nabla_r S_k|^2
+\frac{\lambda_{\sigma}}{2}S_k^4 + (\beta|\phi|^2 -
\frac{1}{2}\lambda_{\sigma}\eta_{\sigma}^2)S_k^2) =
\mathcal{E}_{\sigma} + \frac{\lambda_{\sigma}}{4} \Sigma_4 -k^2
\Sigma \, .
\label{gamma}
\end{equation}
Now to simplify the expression (\ref{gamma}) for $\gamma$, 
we can make use of our
analytic result for $\mathcal{E}_{\sigma}$ (\ref{nham}) to find 
that  $\gamma = -k^2 \Sigma$. We can now rewrite the
eigenvalue equation (following the method used in \cite{LA:CINST}):
\begin{equation}
-\Sigma_2 \nabla_z^2\nu + \gamma\nu + \frac{\lambda_{\sigma}}{2}\Sigma_4[\nu +
\nu^+e^{2ikz}] = \lambda \nu \, .
\label{evalue}
\end{equation}
To simplify this equation further, consider expanding our perturbation
$\nu$
in the form $\nu = e^{ikz}[u_1 +
iu_2]$. Substitution leads to the system of coupled equations:
\begin{equation}
-\Sigma\nabla_z^2u_1 + 2\Sigma k\nabla_zu_2 + (k^2\Sigma + \gamma
+ \lambda_{\sigma}\Sigma_4 - \lambda)u_1 = 0\,,
\label{eq1}
\end{equation}
\begin{equation}
-\Sigma\nabla_z^2u_2 - 2\Sigma k\nabla_zu_1 + (k^2\Sigma + \gamma
- \lambda)u_2 = 0\,.
\label{eq2}
\end{equation}
To solve this system, we expand in Fourier modes, $u_i = a_i e^{ipz}$,
and we use our simple expression for $\gamma$ 
to obtain the following set of linear equations: 
\begin{equation}
(\Sigma p^2 +\lambda_{\sigma} \Sigma_4 - \lambda_p) a_1 +
2i(k.p)\Sigma a_2 = 0\,,
\label{eq3}
\end{equation}
\begin{equation}
(\Sigma p^2 - \lambda_p) a_2 -
2i(k.p)\Sigma a_1 = 0 \, .
\label{eq4}
\end{equation}
The vanishing of the determinant of this system yields our eigenvalues:
\begin{equation}
\lambda_p = p^2 \Sigma + \frac{\lambda_{\sigma}}{2}\Sigma_4 \pm
\sqrt{\left( \frac{\lambda_{\sigma}}{2}\Sigma_4 \right)^2 + 4 \Sigma^2(k.p)^2}
\, .
\label{solevalue}
\end{equation}

Instabilities will occur when one of the possible $\lambda_p$ is
negative, which is equivalent to the condition
\begin{equation}
\Sigma^2 p^4 + 4p^2\Sigma(\frac{\lambda_{\sigma}}{4}\Sigma_4 - k^2
\Sigma) < 0 \, ,
\label{ineq}
\end{equation}
which leads to:
\begin{equation}
\frac{\lambda_{\sigma}}{4}\Sigma_4 - k^2\Sigma < 0
\label{criterion}
\end{equation}
With our ansatz (\ref{Sigmaan}) for the moments $\Sigma_2,\,\Sigma_4$,
 we can use this result to determine 
the critical value of $k_{\rm inst}$ above which a current becomes unstable:
\begin{equation}
k^2_{\rm inst} = \frac{\alpha k_c^2}{\alpha + k_c^2} \, ,
\label{kinst}
\end{equation}
where $\alpha = \frac{\lambda_{\sigma}}{4}
\Sigma_{4o}/\Sigma_o$ and $k_c^2$ is given by
(\ref{c3}). Typically, the coefficient $\alpha$ is of order unity,
and so we can expect $k_{\rm inst}^2 \simeq k^2_c/2$, in good agreement
with numerical estimates. We contrast this quantitatively 
with the heuristic result given in ref.~\cite{TH:SCS} which suggested 
that the onset of instability was indicated by the condition
$\frac{\partial J}{\partial k} <0$; our numerical estimates indicate 
that the precise criterion (\ref{kinst}) is significantly lower.

Note also that through
(\ref{ineq}), we are able to predict the associated wavenumber 
$p$ and hence the typical 
lengthscale of the perturbation for a given unstable current with 
$k > k_{\rm inst}$.  Again, this
is in good agreement with our simulations. (Strictly speaking, a more 
accurate analysis should take into account the effects of 
charge conservation; we will discuss this more elaborate calculation 
elsewhere \cite{LS:DCS}, but we note that it yields the same 
result (\ref{kinst}).)

\begin{figure}
\resizebox{13cm}{17cm}{\includegraphics{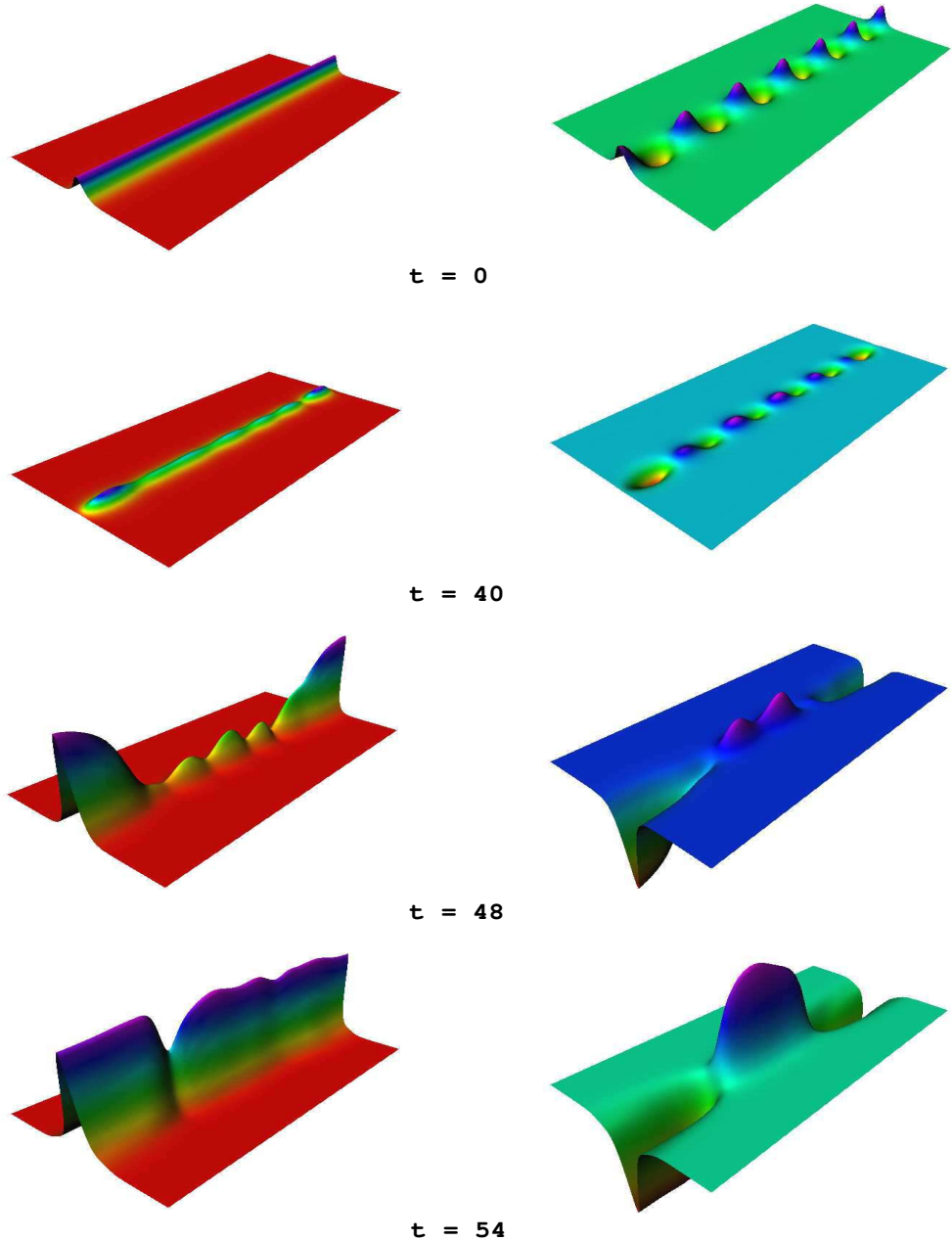}}
\caption{Modulus (left column) and real part (right column) of the
condensate along the string axis at various stages of its evolution, showing
the current
instability. We clearly see that the condensate looses quanta of
winding in this process}
\label{thinst}
\end{figure}

We point out that this analysis has been carried out by looking only
at the condensate part of the energy. Therefore, it is valid in a
broader class of
models than the one we are considering here, since the vortex is somehow
irrelevant. In particular, this instability may appear in any model
exhibiting a second-order $U(1)$ phase transition, and could be
applied in e.g. superconductivity studies of resistive transition.

Our three-dimensional field theory code allows us to follow the 
development and evolution of these
perturbations and their consequences for the structure of the
superconducting string. From fig.~\ref{thinst}, we see that 
the condensate becomes pinched as the instability develops;
if the winding number is sufficiently high, the instability 
will force the condensate down such that $\sigma = 0$ at a localised
point along the string.  This allows the string to lose quanta 
of winding and lowers $k$ in this region.  The condensate then 
bounces back and slowly relaxes into the stable current configuration, 
radiating the energy difference between the initial and the final 
configuration. These numerical results are provided by way of 
illustration but a more extensive study will be reported shortly 
with further details about the simulation code \cite{LS:DCS}. 

\section{Conclusions and prospects}

In this paper we have studied the behaviour and effects 
of the currents and charges on 
superconducting strings and loops. We have derived exact formulae for
the energy of a loop, which exhibits a generic divergence at $\omega^2 - k^2
= 0$, therefore proving that the chiral case is not an attractor, but
rather a repeller. 

Since for stable configurations we typically have $\Sigma
\simeq \Sigma_{QN}\equiv Q/N $, from (\ref{L2}) we can expect the final 
size of the loop
to be very much smaller than its original size (for realistic initial 
conditions). Of course, we have not taken into account the 
vortex-antivortex interactions on scales small compared to the 
string width, so the inevitability of vorton formation is subject 
to this important caveat. In addition we conclude that springs and Q-loops are
not allowed in this theory.

Studying the stability of the current on a straight string, we have
also seen that the superconducting regime is unstable when
the winding is too high. By Lorentz-invariance, it is easy
to see that this will be the case if $m_o^2 v = k^2 - \omega^2 >
k_{\rm inst}^2$, where $m_o^2 \simeq k^2_c \simeq \omega_c^2$. (We believe
that the gauged case will exhibit the same generic features as these 
global currents, though with the small quantitative differences
already discussed.)

Now, from our analysis of the vorton state, we know that
equilibrium typically will be achieved when $\Sigma \simeq
\Sigma_{QN}$. Using our ansatz (\ref{Sigmaf}), this is equivalent to
\begin{equation}
v \simeq \frac{\Sigma_{QN} - \Sigma_o}{\Sigma_{QN} +
\Sigma_o} \, .
\label{conc}
\end{equation}
To ensure stability, we have to impose $v > -k_{\rm inst}^2
/ m_o^2$ which, with (\ref{conc}) and some algebra, leads to the condition
\begin{equation}
\Sigma_{QN} > \frac{m_o^2 - k_{\rm inst}^2}{m_o^2 + k_{\rm inst}^2} \,\Sigma_o
\, .
\label{conc2}
\end{equation}
Loops that do not satisfy (\ref{conc2}) will be unstable and lose
quanta of winding. Hence, $\Sigma_{QN}$ will increase, and the loop
may reach a stable state. This process is associated with energy
radiation, which would be interesting to quantify to determine
possible observational signatures of this phenomenon.

Finally, the perturbative stability analysis carried out for superconducting currents
in the magnetic regime has been extended to the
chiral and electric cases, and appears to establish their stability.
However, this issue is more subtle, and the analysis will be 
published separately \cite{LS:DCS}.

\begin{ack}

Y.L. and E.P.S. gratefully acknowledge very fruitful
conversations with Jose Blanco-Pillado; Y.L. is especially
grateful to him for pointing out ref.~\cite{LA:CINST}. The 
numerical code employed here was originally developed with 
Jonathan Moore \cite{JM:PHD,MSM:CS} to whom we are greatly indebted.

Y.L. is supported by EPSRC, the Cambridge European Trust
and the Cambridge Newton Trust. This work was also supported by PPARC grant no.
PPA/G/O/1999/00603.

Numerical simulations were performed on the COSMOS supercomputer, the
Origin3800 owned by the UK Computational Cosmology Consortium,
supported by Silicon Graphics Computer Systems, HEFCE and PPARC.

\end{ack}

\bibliography{scurrents}
\bibliographystyle{unsrt}

\end{document}